\documentclass[prl,showpacs,twocolumn,amssymb]{revtex4}
\begin{document}

\title{Dynamically Induced Zeeman Effect in Massless QED}
\author{Efrain J. Ferrer}
\author{Vivian de la Incera}
\affiliation{Department of Physics, Western Illinois University,
Macomb, IL 61455, USA}

\begin{abstract}

It is shown that in non-perturbative massless QED an anomalous
magnetic moment is dynamically induced by an applied magnetic field.
The induced magnetic moment produces a Zeeman splitting for
electrons in Landau levels higher than $l=0$. The expressions for the non-perturbative
Lande g-factor and Bohr magneton are obtained. Possible applications
of this effect are outlined.

\end{abstract}
\maketitle The theory of the electron magnetic moment has
historically played an important role in the development of QED. As
it is known, the electron intrinsic magnetic moment
$\overrightarrow{\mu}$ is related to the spin vector
$\overrightarrow{s}$ by
$\overrightarrow{\mu}=g\mu_{B}\overrightarrow{s}$, where
$\mu_{B}=e\hbar / 2mc$ is the Bohr magneton, and $g$ is the Lande
g-factor. One of the great triumphs of the Dirac relativistic theory
for the electron was the prediction of the value $g=2$.
Nevertheless, experimental measurements of the g-factor showed a
deviation from this prediction. The solution of the apparent contradiction came only after
Schwinger calculated the first-order radiative correction to $\overrightarrow{\mu}$, due to
the electron-photon interactions \cite{Schwinger}. Schwinger's results led to
an anomalous magnetic moment with a correction to the g-factor of order
$\frac{g-2}{2}=\frac{\alpha}{2\pi}$)), $\alpha$ being the fine-structure constant.
Subsequently higher-order radiative corrections to g have given rise to a
series in powers of $\alpha/\pi$ \cite{Kinoshita-Lindquist} that is in excellent agreement with the experiment.

Now, in the case of massless QED, one cannot follow Schwinger's approach to obtain
the anomalous magnetic moment. The reason is that an anomalous magnetic moment
would break the chiral symmetry of the massless theory, but this symmetry is protected against
perturbative corrections. However, the chiral symmetry can be broken dynamically
via non-perturbative effects. In fact, such a dynamical symmetry breaking has been shown
to occur if the massless electrons interact with the photons in the presence of a
constant magnetic field. This mechanism is known in the literature \cite{MC}-\cite{Ayala} as the
magnetic catalysis of chiral symmetry breaking ($MC\chi SB$). The phenomenon of $MC\chi SB$
consists of the formation of a chiral condensate due to the dimensional reduction
in the dynamics of the fermions in the Lowest Landau level (LLL). This dimensional reduction makes
the non-perturbative fermion-antifermion interaction
effectively stronger, hence favoring fermion-antifermion pairing even at weak coupling.

All the previous studies of $MC\chi SB$ in QED \cite{MC-QED}-\cite{Ng} focused on the generation of a
fermion dynamical mass. None of them however considered the possibility of a dynamically generated magnetic moment. In
the present paper we are going to show that, along with the dynamical mass, the chiral condensate necessarily produces a
dynamical magnetic moment. Physically it is easy to understand the origin of the new dynamical quantity. The chiral
condensate carries non-zero magnetic moment, since the particles
forming the condensate have opposite spins and opposite charges.
Therefore, chiral condensation will inexorably provide the
quasiparticles with both a dynamical mass and a dynamical magnetic moment.
Symmetry arguments can help us also to better understand this phenomenon. A magnetic moment term
does not break any additional symmetry that has not already been broken by a mass term.
Hence, once $MC\chi SB$ occurs, there is no reason why only one of these parameters should be different from zero.
We will show below that a very important consequence of the dynamically generated
magnetic moment is a splitting in the electron energy
spectrum that can be interpreted as a non-perturbative Zeeman
effect.

To explore the dynamical generation of a magnetic moment in massless
QED, we can start from the Schwinger-Dyson (SD) equation for the
fermion self-energy in the presence of a constant magnetic field
along the $Z$-direction ($F_{12}=H$). We will work in the
quenched-ladder approximation where
\begin{equation}
\label{SD} \Sigma (x,x')=ie^2 \gamma^{\mu}G(x,x')\gamma^{\nu}D_{\mu
\nu}(x-x').
\end{equation}
Here, $\Sigma (x,x')$ is the electron self-energy operator, $D_{\mu \nu}(x-x')$ is the bare photon propagator, and $G(x,x')$ is the full
fermion propagator depending on the dynamically induced quantities and the
magnetic field.

To transform to momentum space in the presence of a magnetic
field we can use the so-called Ritus' method,
originally developed for fermions in \cite{Ritus:1978cj} and later
extended to vector fields in \cite{efi-ext}. In Ritus' approach, the
transformation to momentum space is carried out using the eigenfunctions
$E_{p}^{l}(x)$ of the asymptotic states of the charged
fermions in a uniform magnetic field
\begin{equation}\label{Ep}
\label{E-p} E_{p}^{l}(x)=E_{p}^{+}(x)\Delta(+)+E_{p}^{-}(x)\Delta(-)
\end{equation}
where $\Delta(\pm)=(I\pm i\gamma^{1}\gamma^{2})/2$  are up ($+$) and down ($-$) spin projectors;
$E_{p}^{+/-}(x)=N(l/l-1)exp(p_{0}x^{0}+p_{2}x^{2}+p_{3}x^{3})D_{(l/l-1)}(\rho)$,
with $D_{l}(\rho)$ the parabolic cylinder functions of argument
$\rho=\sqrt{2|eH|}(x_{1}-p_{2}/|eH|)$, and $N(l)=(4\pi|eH|)^{1/4}/\sqrt{l!}$ a normalization constant. The index $l=0,1,2,...$
denotes the Landau levels (LL). The $E_{p}^{l}(x)$ functions
(\ref{Ep}) play the role in the magnetized medium of the usual
plane-wave (Fourier) functions $e^{i px}$ at zero field. They
satisfy the field-dependent eigenvalue equation
\begin{equation}
(\Pi\cdot\gamma)E^{l}_{p}(x)=E^{l}_{p}(x)(\gamma\cdot\overline{p}) \
, \label{eigenproblem}
\end{equation}
with generalized momenta $\Pi _{\mu }=i\partial _{\mu }-eA_{\mu }$
and $\overline{p}=(p_{0},0,-sgn(eH) \sqrt{2|eH|l},p_{3})$.

In momentum space the fermion self energy is given by
\begin{eqnarray}\label{P-Self-Energy}
\Sigma(p,p')= \int dxdy
\overline{E}_{p}^{l}(x)\Sigma(x,y)E_{p}^{l}(y) \nonumber
\\
=(2\pi)^4\widehat{\delta}^{(4)}(p-p')\Pi(l)\widetilde{\Sigma}^l
(\overline{p})
\end{eqnarray}
since the $E_{p}^{l}$ are precisely linear combinations of
the eigenfunctions of the fermion self energy in the presence of a
magnetic field \cite{Ritus:1978cj}. In (\ref{P-Self-Energy})
$\overline{E}_{p}^{l}\equiv \gamma^{0} (E_{p}^{l})^{\dag}\gamma^{0}$, and we used that $\int d^{4}x \overline{E}_{p}^{l}(x)E_{p'}^{l'}(x)=(2\pi)^4 \widehat{\delta}^{(4)}(p-p')\Pi(l)$ with $\widehat{\delta}^{(4)}(p-p')=\delta^{ll'} \delta(p_{0}-p'_{0}) \delta(p_{2}-p'_{2}) \delta(p_{3}-p'_{3})$
and $\Pi(l)=\Delta(+)\delta^{l0}+I(1-\delta^{l0})$ \cite{Wang}.

As proven in \cite{WT-identity}, in the presence of a magnetic field $H$, the general structure of
$\widetilde{\Sigma}^{l} (\overline{p})$ consistent with
the Ward-Takahashi identity in the ladder approximation is
\begin{eqnarray}\label{Sigma}
\widetilde{\Sigma}^l
(\overline{p})&=& Z_{\|}^{l}(\overline{p},F)\gamma\cdot
\overline{p}_{\|}+Z_{\bot}^{l}(\overline{p},F)\gamma\cdot
\overline{p}_{\bot}  \nonumber
\\
&+& M^{l}(\overline{p},F)I+T^{l}(\overline{p},F)\overline{F}^{\mu
\nu}\sigma_{\mu \nu} \qquad
\end{eqnarray}
Here, $\overline{F}^{\mu \nu}=F^{\mu \nu}/|H|$,
$\overline{p}_\mu^{\bot}=(0,0,-sgn(eH) \sqrt{2|eH|l},0)$ and
$\overline{p}_\mu^{\|}=(p_{0},0,0,p_{3})$. The coefficients $M^{l},
Z^{l}$, and $T^{l}$ depend on the field strength $F$, LL $l$
and momentum $\overline{p}$. $M^{l}$ is the dynamical mass already
considered in previous works on $MC\chi SB$ \cite{MC}-\cite{Ayala}. $T^{l}$
corresponds to the dynamically induced magnetic moment and should be
found, along with $M^{l}$, from the SD equations. The operator
$\widetilde{\Sigma}^l (\overline{p})$ can be conveniently written, with the help of the projectors
\begin{equation}  \label{projector}
\Lambda^{\pm}_{\|}=\frac{1}{2}(1\pm\frac{\gamma^{\|}\cdot
\overline{p}_{\|}}{|\overline{p}_{\|}|}), \qquad
\Lambda^{\pm}_{\bot}=\frac{1}{2}(1\pm i\gamma^2),
\end{equation}
as
\begin{eqnarray}\label{Sigma-2}
\widetilde{\Sigma}^l
(\overline{p})&=&Z_{\|}^{l}(\Lambda^{+}_{\|}-\Lambda^{-}_{\|})|\overline{p}_{\|}|
+iZ_{\bot}^{l}(\Lambda^{-}_{\bot}-\Lambda^{+}_{\bot})|\overline{p}_{\bot}|
\nonumber
\\
&+&(M^{l}+T^{l})\Delta(+)+(M^{l}-T^{l})\Delta(-)
\end{eqnarray}

Using the $E^{l}_{p}$ transformation, the full fermion propagator in momentum
space is given by
\begin{eqnarray}\label{Propagator}
G^{l}(p-p')&=& \int dxdy \overline{E}_{p}^{l}(x)G(x,y)E_{p'}^{l'}(y)\nonumber
\\
&=&(2\pi)^4\widehat{\delta}^{(4)}(p-p')\Pi(l)\widetilde{G}^{l}(\overline{p})
\end{eqnarray}
where
\begin{eqnarray*}
\widetilde{G}^{l}(\overline{p})=\frac{1}{\gamma\cdot
\overline{p}-\widetilde{\Sigma}^l(\overline{p})}=
\end{eqnarray*}
\begin{eqnarray}\label{P-Propagator}
&=&\frac{N^l(T,V_{\|})}{D^l(T)}\Delta(+)\Lambda^{+}_{\|}
+\frac{N^l(T,-V_{\|})}{D^l(-T)}\Delta(+)\Lambda^{-}_{\|}\nonumber
\\
&+&\frac{N^l(-T,V_{\|})}{D^l(-T)}\Delta(-)\Lambda^{+}_{\|}
+\frac{N^l(-T,-V_{\|})}{D^l(T)}\Delta(-)\Lambda^{-}_{\|}\nonumber
\\
&-&iV_{\bot}^l(\Lambda^{+}_{\bot}-\Lambda^{-}_{\bot})[\frac{\Lambda^{+}_{\|}\Delta(+)+\Lambda^{-}_{\|}\Delta(-)}{D^l(T)}\nonumber
\\
&+&\frac{\Lambda^{-}_{\|}\Delta(+)+ \Lambda^{+}_{\|}\Delta(-)}{D^l(-T)}]
\end{eqnarray}
with coefficients
\begin{eqnarray}\label{Coefficients}
N^l(T,V_{\|})&\equiv&M^l-T^l-V_{\|}^l \nonumber
\\
D^l(T)&\equiv&(M^l)^2-(V_{\|}^l+T^l)^2+(V_{\bot}^l)^2 \nonumber
\\
V_{\|}^l&\equiv&(1-Z_{\|}^{l})|\overline{p}_{\|}|\nonumber
\\
V_{\bot}^l&\equiv&(1-Z_{\bot}^{l})|\overline{p}_{\bot}|=(1-Z_{\bot}^{l})\sqrt{2|eH|l}
\end{eqnarray}

Transforming Eq.(\ref{SD}) to momentum space with the help of the $E_{p}^{l}$ functions, taking the photon propagator in the Feynman gauge, $D_{\mu \nu}(x-x')=\int
\frac{d^4q}{(2\pi)^4} \frac{e^{iq\cdot (x-x')}
}{q^{2}-i\varepsilon}g_{\mu \nu}$, and carrying out derivations and approximations
similar to those done in \cite{Ng}, we obtain that the SD equation for arbitrary
Landau level $l$ is given by
\begin{eqnarray}\label{SD-arbitrary-LL}
\widetilde{\Sigma}^{l}(\overline{p})\Pi(l)&=&
ie^2(2eH)\Pi(l)\int\frac{d^4\widehat{q}}{(2\pi)^4}
\frac{e^{-\widehat{q}^2_\bot}}{\widehat{q}^2}[\gamma_{\mu}^{\|}\widetilde{G}^{l}(\overline{p-q})\gamma_{\mu}^{\|}\nonumber
\\
&+&\Delta(+)\gamma_{\mu}^{\bot}\widetilde{G}^{l+1}(\overline{p-q})\gamma_{\mu}^{\bot}\Delta(+)\nonumber
\\
&+&\Delta(-)\gamma_{\mu}^{\bot}\widetilde{G}^{l-1}(\overline{p-q})\gamma_{\mu}^{\bot}\Delta(-)]
\end{eqnarray}
where $\overline{p-q} \equiv(p_{0}-q_{0},0,-sgn(eH)
\sqrt{2|eH|n},p_{3}-q_{3})$ for $n=l-1,l,l+1$ and the normalized quantities are defined as
$\widehat{Q}_{\mu}=Q_{\mu}/\sqrt{2|eH}$. Since the equation for a given
Landau level $l$ involves dynamical parameters that depend on $l$,
$l-1$ and $l+1$, the SD equations for all the LL's actually form a
system of infinite coupled equations. Fortunately, in the
infrared region, the leading contribution to each equation will come
from the propagators with the lower LL's, since the magnetic field
appearing in the denominator of the fermion propagator for $l\neq 0$
acts as a suppressing factor. Using this approximation, one can find
a consistent solution at each level. On the other hand, the
solutions for any $M^{l}$ and $T^{l}$ can be ultimately expressed in
terms of the LLL solution, indicating that the physical origin of
all the dynamical quantities is actually due to the infrared
dynamics taking place at the LLL.
For the LLL (l=0) case, the leading contribution to the RHS of (\ref{SD-arbitrary-LL}) comes from the $\widetilde{G}^{0}(\overline{p-q})$ term, and we find
\begin{eqnarray}\label{P-SD-2}
(M^0+T^0)+Z_{\|}^0(\Lambda^{+}_{\|}-\Lambda^{-}_{\|})|\overline{p}_{\|}|=ie^{2}(2|eH|)
\nonumber
\\
\cdot \int
\frac{d^4q}{(2\pi)^4}\frac{e^{-\widehat{q}^{2}_{\bot}}}{\widehat{q}^2}
\frac{(M^0+T^0)}{(\overline{p}_{\|}-\overline{q}_{\|})^{2}-(M^0+T^0)^2}
\end{eqnarray}

Eq. (\ref{P-SD-2}) implies that $Z_{\|}^0 =0$; while for the
combination $M^0+T^0$ it gives, in the infrared limit ($p_{\|}\sim 0$),
\begin{equation}\label{Mass-Eq-2}
1=ie^2(4|eH|)\int\frac{d^4\widehat{q}}{(2\pi)^4}
\frac{e^{-\widehat{q}^2_\bot}}{\widehat{q}^2}\frac{1}{(M^{0}+T^{0})^2-
q_{\|}^2}
\end{equation}
If $M^{0}+T^{0}$ is replaced in (\ref{Mass-Eq-2}) by the dynamical mass $m_{dyn}$ of Refs.
\cite{MC-QED}-\cite{Ng}, Eq. (\ref{Mass-Eq-2}) turns identical to the
gap equation found there. Hence, the solution of (\ref{Mass-Eq-2})
is formally the same as the one found in \cite{MC-QED}-\cite{Ng}, but with the combination $M^{0}+T^{0}$
now playing the role previously played only by the dynamical mass. Hence,
\begin{equation}
\label{Mass-Eq-Solution} M^0+T^0\simeq \sqrt{2|eH|}
e^{-\sqrt{\frac{\pi}{\alpha}}},
\end{equation}
As in \cite{MC-QED}-\cite{Ng}, this solution is obtained considering that $M^{0}+T^{0}$  does not
depend on the momentum, an assumption consistent within the ladder approximation \cite{BCMA}.
As proved in \cite{Improved-LA}, when the  polarization effect was included in the gap equation through
the improved-ladder approximation, the solution for $m_{dyn}$ was of the same form as (\ref{Mass-Eq-Solution}), but with the
replacement $\sqrt{\pi/\alpha} \rightarrow \pi/\alpha \log (\pi/ \alpha)$ in the exponent. Since the inclusion of the magnetic moment in the LLL SD equation
merely implies the replacement $m_{dyn}\rightarrow M^{0}+T^{0}$, it is expected that a similar effect will occur in the solution (\ref{Mass-Eq-Solution}).
However, this effect will not qualitatively change the nature of our findings.

Since in the LLL propagator $G^{0}(p-p')$ the dynamical mass $M^{0}$ and magnetic moment $T^{0}$ always enter
through the combination $M^{0}+T^{0}$, the solution of the LLL SD equation (\ref{Mass-Eq-Solution}) can only determine
the sum of these dynamical parameters. This indicates that at the LLL, the effect of a magnetic moment is irrelevant,
it just redefines the rest energy due to the replacement $m_{dyn}\rightarrow M^{0}+T^{0}$. This is physically natural, since
the electrons in the LLL can only have one spin projection, so for them there is no spin degeneracy
and hence, no possible energy splitting due to the magnetic moment. $E^0=M^0+T^0$ represents then
a dynamically induced rest-energy. This can be easily seen considering the Dirac equation for the electrons in the LLL with dynamically induced parameters,
\begin{equation}
\label{LLL-electron-eq}
[\overline{p}^{\|}\cdot\widetilde{\gamma}_{\|}-E^0]\psi_{LLL} =0,
\end{equation}
where $\psi_{LLL}$ is the spin-up two-component wave-function. Eq. (\ref{LLL-electron-eq}) coincides with the
the free (1+1)-Thirring model \cite{Th-M}, with corresponding
gamma matrices $\widetilde{\gamma}_{0}=\sigma_1$,
$\widetilde{\gamma}_{3}=-i\sigma_2$, where $\sigma_i$ are the Pauli
matrices. The dispersion relation of the electrons in the LLL
obtained from (\ref{LLL-electron-eq}), $p_0^2=p_3^2+(E^0)^2$,
is in agreement with the above discussion. As we will see below, the interesting effect
associated to $T$ comes from the higher $LL's$.

For electrons in the first LL ($l=1$), the leading contribution to the RHS of (\ref{SD-arbitrary-LL}) in the infrared limit ($p_{\|}\sim
0$) comes from the term containing $\widetilde{G}^{0}(\overline{p-q})$. Then,
\begin{eqnarray}\label{SD-1-LL}
Z_{\bot}^{(1)}\gamma_2(2|eH|)+(M^{1}+T^{1})\Delta(+)+(M^{1}-T^{1})\Delta(-)
= \nonumber
\\
=ie^2(4|eH|)\Delta(-)\int\frac{d^4\widehat{q}}{(2\pi)^4}
\frac{e^{-\widehat{q}^2_\bot}}{\widehat{q}^2}\frac{E^{0}}{(E^{0})^2-q_{\|}^2}\qquad\qquad
\end{eqnarray}
From (\ref{SD-1-LL}) we obtain the solutions
\begin{equation}\label{M}
M^{1}=-\label{T} T^{1}=\frac{1}{2} E^{0}=\sqrt{|eH|/2}
e^{-\sqrt{\frac{\pi}{\alpha}}},\quad Z_{\bot}^{1}=0
\end{equation}
This result corroborates the relevance of the LLL dynamics (both $M^{1}$ and $T^{1}$ are determined by $E^{0}$)
in the generation of the dynamical mass and magnetic moment for electrons in the first LL.
Given that the magnitude of the magnetic moment for the electrons in the first LL
is determined by the dynamically generated rest-energy of the electrons in the LLL, any modification of the theory
producing an increase in $E^{0}$ will, in turn, drives an increase in the magnitude of $T^{1}$.
From the experience with the $MC\chi SB$ phenomenon, such modifications could be for example, lowering the space dimension
\cite{Gusynin}, introducing scalar-fermion interactions
\cite{Vivian, BCMA}, or considering a non-zero bare mass
\cite{Mass}.

Let us find now the dispersion relations for electrons in higher
LL's, taking into account the dynamically induced quantities. Starting from the modified electron
equation in the presence of the magnetic field
\begin{equation}
\label{electron-eq}
[\overline{p}\cdot\gamma-M^lI-iT^l\gamma^1\gamma^2]\psi_l =0,
\end{equation}
the dispersion relations are found from
\begin{eqnarray}
\label{determinant}
&det&[\overline{p}\cdot\gamma-M^lI-iT^l\gamma^1\gamma^2]=
\nonumber
\\
&=&[(M^l)^2-(\overline{p}_{\|}-T^l)^2+\overline{p}_{\bot}^2]
\nonumber
\\
&\times&[(M^l)^2-(\overline{p}_{\|}+T^l)^2+\overline{p}_{\bot}^2]
=0.
\end{eqnarray}
yielding
\begin{equation}
\label{disp-relat-1} p_{0}^{2}=p_{3}^{2}+[\sqrt{(M^l)^{2}+2eHl}\pm
T^l]^2,
\end{equation}
and thus showing that the induced magnetic moment breaks the energy degeneracy between the
spin states in the same LL.

In particular for $l=1$, plugging (\ref{M}) into
(\ref{disp-relat-1}), taking into account that
$\widehat{M}^1, \widehat{T}^1 \ll 1$, and Taylor expanding the term in parenthesis,
the dispersion relations can be expressed as
\begin{equation}\label{disp-relat-3}
p_{0}^{2}\simeq p_{3}^{2}+2eH+(M^1)^2+(T^1)^2\pm 2T^1\sqrt{2eH},
\end{equation}
thereby producing an energy splitting
\begin{equation}\label{energy-splitting} \Delta E=|2T^{1}|=2
\sqrt{|eH|/2} e^{-\sqrt{\pi /\alpha}}
\end{equation}

Expression (\ref{energy-splitting}) can be conveniently written in the well known form of
the Zeeman energy splitting for the two spin projections
\begin{equation}\label{Zeeman-splitting}
\Delta E=\widetilde{g}\widetilde{\mu}_B H
\end{equation}
where $\widetilde{g}$ and $\widetilde{\mu}_B$ are the
non-perturbative Lande g-factor and Bohr magneton given respectively by
\begin{equation}\label{Zeeman-splitting-const}
\widetilde{g}=2e^{-2\sqrt{\pi /\alpha}},\quad \widetilde{\mu}_B
=\frac{e}{2M^1}
\end{equation}
Notice that the Lande g-factor depends non-perturbatively on the
coupling constant $\alpha$, and that the Bohr magneton is given in
terms of the dynamically induced electron mass.

We want to call attention to possible applications of the dynamically
induced Zeeman effect obtained in this paper. One area of potential interest is condensed
matter, since recent experiments \cite{graphene} have shown that the
2-dimensional crystalline form of carbon, known as graphene, has
charge carriers that behaves as massless Dirac electrons. In
particular, a phenomenon where the dynamically induced Zeeman effect
can bring some new light is the lifting of the fourfold degeneracy
of the $l=0$ LL, and twofold degeneracy of the $l=1$ LL in the
recently found quantum Hall states corresponding to filling factors
$\nu =0,\pm 1, \pm 4$ under strong magnetic fields \cite{QH}. Notice
that dispersion relations similar to (\ref{disp-relat-1}) were found
within certain region of the parameter space in a 2-dimensional
modeling of Dirac quasiparticles in graphene with magnetically
catalyzed masses and other order parameters connected to quantum
Hall ferromagnetism \cite{Gorbar}.

Another domain where the finding we are reporting can be of interest
is color superconductivity. An important aspect of color
superconductivity is its magnetic properties
\cite{alf-raj-wil-99/537}-\cite{Vortex}. In spin-zero color
superconductivity, although the color condensate has non-zero
electric charge, there is a linear combination of the photon and a
gluon that remains massless, hence giving rise, in both the 2SC and
CFL phases, to a long-range remnant "rotated-electromagnetic" field
\cite{alf-raj-wil-99/537}. To understand this, notice that, the
quarks participating in the pairing are neutral or have equal and
opposite "rotated" $\widetilde{Q}$-charge. That is, the condensate
is always $\widetilde{Q}$-neutral. In the case that the pair is
formed by $\widetilde{Q}$-charged quarks of opposite sign, although
the condensate is $\widetilde{Q}$-neutral, an applied magnetic field
can interact with the quarks forming the pair \cite{MCFL}. Hence,
with respect to the "rotated-electromagnetism" the
color-superconducting pair resembles the chiral condensate under a
conventional electromagnetic field. It should be expected then that
a non-perturbative Zeeman effect can also be induced in a color
superconductor under an applied magnetic field. Since, on the other hand,
the Meissner instabilities that appear in some density regions of the color superconductor can be removed by
the induction of a magnetic field \cite{Vortex}, it will be
interesting to investigate what could be the role in this process of
a dynamically induced magnetic moment.

{\bf Acknowledgments:} We thanks V.P. Gusynin, V.A. Miransky and I.A. Shovkovy for comments. This work has been supported in part by DOE Nuclear Theory grant DE-FG02-07ER41458.

\end{document}